# Functional single-layer graphene sheets from aromatic monolayers

Dan G. Matei[1], Nils-Eike Weber[1], Simon Kurasch[3], Stefan Wundrack[2], Mirosław Woszczyna[2], Miriam Grothe[2], Thomas Weimann[2], Franz Ahlers[2], Rainer Stosch[2], Ute Kaiser[3], and Andrey Turchanin[1*]

[1]Faculty of Physics, University of Bielefeld, 33615 Bielefeld, Germany
[2]Physikalisch-Technische Bundesanstalt, 38116 Braunschweig, Germany
[3]Electron Microscopy Department of Material Sciences,
University of Ulm, 89081 Ulm, Germany

e-mail: turchanin@physik.uni-bielefeld.de
Tel.: +49-521-1065376
Fax: +49-521-1066002

**Abstract**

We demonstrate how self-assembled monolayers of aromatic molecules on copper substrates can be converted into high-quality single-layer graphene using low-energy electron irradiation and subsequent annealing. We characterize this two-dimensional solid state transformation on the atomic scale and study the physical and chemical properties of the formed graphene sheets by complementary microscopic and spectroscopic techniques and by electrical transport measurements. As substrates we successfully use Cu(111) single crystals and the technologically relevant polycrystalline copper foils.

**TOC Graphic**

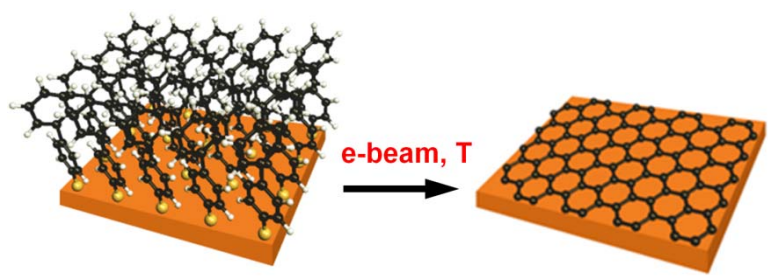

After the first experimental studies of the exciting electronic properties of individual graphene sheets[1-3], the research landscape in physics, chemistry and materials science has been strongly dominated by two-dimensional (2D) carbon materials in particular due to their promises for novel applications in nanotechnology.[4-6] Despite recent progress in fabrication of graphene by various techniques (e.g., via chemical exfoliation of graphite[7], CVD growth on metals[8] or thermal graphitization of silicon carbide[9]), technologically efficient tailoring of this truly 2D material with specific application-dependent properties is still a demanding task. The main challenges include large-scale fabrication of homogeneous graphene sheets with well-controlled thickness and crystallinity, chemical functionalization of graphene without impairing the electronic structure, direct graphene growth on technologically relevant substrates, fabrication of functional graphene nanostructures and lowering the production costs.[10] Methods towards graphene based on molecular self-assembly[11,12] possess high potential for addressing these challenges, however, they have been comparably little investigated.[13-16] Here we demonstrate how self-assembled monolayers (SAMs) of aromatic molecules on copper substrates can be converted into monolayer graphene using low-energy electron irradiation (50 eV) and subsequent annealing (~800 °C). We characterize this 2D solid state reaction on the atomic scale and study the physical and chemical properties of the formed graphene sheets by complementary microscopic and spectroscopic techniques and by electrical transport measurements. As substrates we successfully used Cu(111) single crystals and the technologically relevant polycrystalline copper foils. Because SAMs can easily be prepared on chemically diverse substrates (metals, semiconductors, insulators) of various sizes and shapes[17], and the areas converted into graphene are simply defined by the electron irradiated regions, we expect that our findings will strongly facilitate the fabrication of graphene with tunable properties both for the wafer-scale and for the nanoscale applications using defocused and focused electron beams, respectively.

Our route to graphene from organic *self-assembled monolayers*[17] (SAMs) is schematically presented in Figure 1. It consists of three consecutive production steps: (i) formation of an aromatic SAM with a well-defined surface density of the carbon atoms on a solid substrate; (ii) electron-irradiation-induced crosslinking of the SAM into a dielectric *carbon nanomembrane* (CNM) with high thermal stability[18]; (iii) temperature-induced conversion of CNM into *graphene* via annealing in vacuum or under protective atmosphere. In the following we characterize in detail each of the steps of this conversion on catalytically active copper substrates, and we discuss advantages of our approach for technological applications of graphene.

The first step in the fabrication of graphene sheets from organic monolayers is the self-assembly of aromatic molecules on a solid substrate, Figures 1a, b. To this end, we employed vacuum vapor deposition of 1,1'-biphenyl-4-thiols (BPTs) on the atomically clean copper substrates at room temperature (details in Supporting Information (SI)). Figure 2a presents a scanning tunneling microscopy (STM) image of a BPT SAM on the Cu(111) single crystal surface directly after vapor deposition. As can be seen, the BPT molecules form a highly ordered monolayer on Cu(111) that exhibits various rotational domains of the same densely packed structure. One of these domains, imaged by high-resolution STM is presented as an inset in Figure 2a (see also SI Figure 1). Low energy electron diffraction (LEED) patterns obtained from the BPT SAM, like the one displayed in Figure 2a, indicate that the molecular structure is incommensurate to the substrate. An approximate unit cell that satisfies both the STM and LEED data is characterized by the vectors with the lengths of 5.00 Å and 5.35 Å and an angle of 122.5°. This unit cell is rotated with respect to Cu(111) by an angle of 16.5° (see SI Figure 2a) and shows twelve rotational domains due to the hexagonal symmetry of the substrate. In agreement with the STM and LEED results the formation of a densely packed BPT SAM is confirmed by X-ray photoelectron spectroscopy (XPS) measurements, shown in Figure 3a. The C1s signal has a binding energy (BE) of 284.6 eV (green) with a full width at half maximum (FWHM) of 1.1 eV and is accompanied by a shoulder at 285.5 eV (red) due to C-S bonds. The S2p signal consists of two doublets demonstrating the presence of two sulfur species on Cu(111) with a branching ratio between the $S2p_{3/2}$ and $S2p_{1/2}$ components of 2:1 due to the spin-orbit coupling. A species with the BE of the $S2p_{3/2}$ component at 162.7 eV (red) contributes to ~80% of the total intensity showing the formation of thiolates[19]. The second doublet with the lower BE of 161.2 eV (blue) is characteristic for the formation of copper sulfides[20], which may result from the partial decomposition of BPT molecules during their vapor deposition on the reactive copper substrate (see also discussion below). Note that this additional sulfur species does not impair the self-assembly of the intact BPT molecules on Cu(111) (see Figure 2a), which is reflected in a much higher structural quality of the formed SAM in comparison to the BPT SAM on Au(111)[21], although on the latter surface only thiolate species are observed by XPS.[21] An effective thickness of the monolayer obtained from the attenuation of the Cu2p signal is ~9 Å, which corresponds well to a nearly vertical arrangement of BPT molecules in the SAM.

The next step in the fabrication of graphene is the electron-irradiation-induced crosslinking of the BPT SAM[22,23] resulting in the dehydrogenation of BPT molecules and formation of the CNM (see Figure 1c) with an extremely high thermal stability[23,25]. An STM image in Fig. 2b shows topographical changes induced in the BPT SAM on Cu(111) (compare with Figure 2a) via irradiation with 50 eV electrons and a dose of 50 mC/cm$^2$ (~3000 electrons per 1 nm$^2$). To a large extent the initially well-ordered alignment of the BPT SAM is lost. As obtained from the STM scans, the root-mean-square (RMS) roughness of the surface increases from 0.3 Å for the BPT SAM to more than 1 Å for the CNM. Thereby, the long-range order disappears, evidenced by vanishing LEED patterns from the sample. However, a short-range order persists across the surface. Periodicities of 5.5 Å are detected by STM resembling the structural features in the pristine SAM (see inset in Figure 2b). Upon crosslinking the amount of copper sulfides increases to ~60 %, as observed from the XPS measurements (see Figure 3b), indicating the decomposition of the C-S bonds. Also the C1s signal experiences changes in accord with the introduced structural modifications[23]. Thus, the BE of the main peak shifts to 284.3 eV and the FWHM increases by 0.1 eV (1.2 eV) in comparison to the pristine BPT SAM, whereas the intensity decreases by ~5% showing some desorption of carbon; the shoulder due to the C-S species shifts to 285.2 eV and its intensity increases by ~30 %.

Finally, Figures 2c-e show the conversion of the CNM (Figure 2b) into graphene upon annealing in vacuum. The presented STM and LEED measurements were conducted at room temperature after the preceding annealing steps. Figure 2c demonstrates an intermediate stage of this 2D solid state transformation after annealing the sample for 15 minutes at 730 °C. It can be seen by STM that most of the surface is still rough showing the same structure as for the non-annealed sample (Figure 2b). However, a few islands with flat areas, indicated with arrows, have been formed. In comparison to the rest of the surface their corrugation is very low with a RMS value of only 0.4 Å. High-resolution STM imaging shows a regular hexagonal structure within these areas, which results in the formation of a LEED pattern (see inset of Figure 2c). From STM and LEED data we conclude that the unit cell of this structure has a length of 6.75 Å and a rotational angle with respect to the substrate of 19.1° (see SI Figure 2b). As seen from the insets of Figure 2c, the simulated LEED pattern matches the experimental one very well. We assign the observed structure to the formation of a superstructure between the graphene lattice and the Cu(111) substrate. Such a phenomenon, often referred to as the formation of moiré patterns, is routinely observed in the growth of graphene on metal substrates[26,27]. In this example the superstructure has a unit cell that is only about three times larger than the unit cell of graphene, making it difficult to directly resolve the atomic structure in STM.

The STM and LEED data for the complete conversion of the CNM into graphene are presented in Fig. 2d. Here, the same sample as in Figure 2c was subsequently annealed for two hours at 800 °C. This treatment causes a drastic change in the topology. A very smooth surface has been formed across the sample resulting in the appearance of a new LEED pattern. High-resolution STM imaging reveals the presence of a hexagonal structure on this surface (see inset of Figure 2d); by increasing the resolution, the honeycomb lattice of a graphene monolayer is clearly imaged by STM (see Figure 4e). Its lattice is rotated with respect to the Cu(111) substrate by ~38°, resulting in the formation of a hexagonal superstructure. In comparison to the superstructure presented in Figure 2c, its unit cell is much larger and has a lattice constant of 2.2 nm and a rotational angle of 23.4° with respect to the substrate (see SI Figure 2c). The simulated LEED pattern, presented in the inset of Figure 2d, reflects most but not all experimentally observed diffraction spots, indicating that also other superstructures with different lattice constants can be present. The hexagonal superstructure discussed in this paragraph was also occasionally imaged by STM after annealing at 730 °C, but its surface density was not sufficient to contribute to the LEED pattern. These observations strongly suggest that during the nucleation and growth of graphene, the crystallites may undergo structural reorientations with respect to the substrate.

Figure 3c shows the XPS spectra for the CNM sample after its complete conversion into graphene (see Figures 2d-e). As a result of this conversion, the FWHM of the C1s signal (BE = 284.5 eV) significantly decreases to a value of 0.9 eV, which corresponds to the resolution of our spectrometer. The signal intensity decreases to ~70% of the initial value for a pristine BPT SAM showing desorption of carbon from the CNM upon the conversion into graphene. Also the S2p signal experiences significant changes, its intensity reduces to ~60% of the initial value and the shape shows only the presence of a copper sulfide species (BE $Sp_{3/2}$ = 161.5 eV). The desorption of this species from the copper substrate is hindered by the intrinsic stability of copper sulfides[20] and by the presence of the graphene layer, which acts as a diffusion barrier for sulfur atoms. Even much longer annealing times (~12 h) do not reduce the intensity of the copper sulfide peak substantially.

From the surface density of the BPT SAM on Cu(111) obtained by STM and LEED and taking into account desorption of carbon during the crosslinking and annealing, the thickness of the graphene can be estimated (see SI p.6). The calculation shows that after the conversion precisely *a single-layer of graphene* is formed on the copper substrate. To further support this result, we characterized the graphene formed on Cu(111) by Raman spectroscopy. A typical Raman spectrum is shown in Fig. 2f. The G- and 2D-peaks are located at 1583 $cm^{-1}$ and 2672 $cm^{-1}$. The Lorentzian-shape of the 2D peak with the FWHM of 37 $cm^{-1}$, in combination with the low-intensity D-peak at 1340 $cm^{-1}$, clearly confirm the formation of single-layer graphene with high structural quality. Note that annealing of the pristine (non-cross-linked) BPT SAM results in desorption of the monolayer at temperatures above 120 °C. After annealing at ~800 °C only a small amount of the sulfide species are detected by XPS on the copper substrate. Thus, the high thermal stability of CNMs is of key importance for their conversion into graphene via annealing.

All experiments described in the previous paragraphs were conducted inside an ultra-high vacuum chamber, except Raman spectroscopy, employing a Cu(111) single crystal as a substrate. In the following we demonstrate that high quality graphene can also be grown on copper foils, which is relevant for a wide spectrum of technological applications. Graphene sheets grown on copper foils were then transferred[11,28] onto Si-wafers with a 300 nm thick $SiO_2$ layer for Raman spectroscopy, XPS, and electrical transport measurements, or onto transmission electron microscope (TEM) grids for characterization with a 80 kV aberration corrected TEM (AC-TEM). The Raman spectroscopy data (see Figure 4a) show an evolution of the D, G and 2D peaks as a function of temperature. A gradual conversion of a CNM into graphene with temperature is clearly seen from these data. For the highest annealing temperature (830 °C) the same features as known for single-layer graphene prepared by mechanical exfoliation with the G-peak at 1587 $cm^{-1}$ and narrow Lorentzian-shape 2D-peak at 2680 $cm^{-1}$ (FWHM=24 $cm^{-1}$)[29] are observed after the conversion. The low-intensity D-peak at 1342 $cm^{-1}$ indicates defects, which may result from graphene grain boundaries observed by STM (see SI Figure 4) and TEM (see following). Complementary to these results, XPS shows that the graphene transferred onto a silicon wafer consists only of carbon species, Figure 3d. The sulfur species detected directly after the growth on copper substrates (see Figure 3c) is no longer detected, confirming its assignment to copper sulfides.

To further characterize the grown graphene sheets we employed AC-TEM, which proved to be an extremely powerful tool for investigating the structure of graphene from the micron (via dark-field imaging[30]) down to the single atom scale (via high-resolution (HR)-TEM imaging[31]) scale. Figure 4b and 4c present an AC-HRTEM image of the suspended graphene sheet at 80 kV and a diffraction pattern obtained from a much wider region than shown in Figure 4b, respectively. The intensity distribution of the diffraction pattern unambiguously proves the single-layer nature of the sheet.[32] Complementary to the STM on Cu(111) (see Fig. 2e), which probes the electron charge density, in TEM the atomic potential is probed showing both in imaging (Figure 4b) and in diffraction (Figure 4c) that atomically perfect single-layer graphene has been formed. Hence, we have demonstrated a complete conversion of the CNM into a monolayer of graphene upon annealing also on copper foils. The polycrystallinity of the sample can be evaluated from the low magnification dark-field TEM images shown in Figure 4d. Different colors correspond to different in-plane lattice orientations of the graphene crystallites (see also SI Figure 5 and video file). The histogram of the grain size (Figure 4e) reveals graphene crystallites with sizes up to ~1.2 $\mu$m and with a mean size of ~300 nm.

The electrical transport properties of the graphene films synthesized on copper foils were studied by four point measurements in the Hall bar geometry (see inset in Figure 4g). Figure 4f presents the ambipolar electric field effect, which was observed in the samples. The room temperature charge carrier mobility, extracted from the data at a hole concentration of $1 \times 10^{12}$ cm$^{-2}$, has a high value of ~1600 cm$^2$/Vs. We further characterized the transport properties at low temperatures (T = 0.3 K) in a magnetic field of 15 T. By varying the charge density with the back-gate voltage of the devices, Shubnikov - de Haas oscillations and resistivity plateaus of the quantum Hall effect specific for a single-layer graphene[2] were clearly observed, Fig. 4h. These results unambiguously confirm the high electronic quality of the grown graphene single-layers making them attractive for applications.

In summary, we have demonstrated that self-assembled monolayers of biphenylthiols on Cu(111) single crystals and copper foils can be converted into high quality graphene with attractive technological properties simply via the electron irradiation and subsequent annealing. This 2D solid state reaction can be tuned by temperature, which enables adjusting the crystallinity of the formed graphene monolayers. Since only the electron-beam irradiated areas undergo this conversion, we expect that both large-area graphene sheets and graphene nanostructures of various architectures (e.g., nano-ribbon, dot, anti-dot patterns) can be generated from SAMs employing either electron flood guns (as in this work) or focused electron beams, respectively. The lateral resolution of the generated nanostructures is defined by the resolution of electron-beam lithography, which has been shown to be 7 nm for SAMs[33]. Although this study addresses one molecular precursor only, diverse aromatic molecules[18] can be used for the described route as well. In this way tuning the thickness of graphene layers or introducing well-defined concentrations of dopants by employing dopant-containing molecules should now be possible. Moreover, molecular self-assembly can be conducted on non-planar surfaces, thus it is also feasible to create graphene structures of any three-dimensional shape. Since SAMs can be formed also on insulating substrates, it is promising to directly grow graphene sheets on these substrates for use in electronic or optical devices. We expect that our route to graphene from well-defined aromatic SAMs will strongly facilitate applications of this material in nanotechnology.

**Acknowledgements**

D.G.M., N.-E.W. and A.T. thank the Deutsche Forschungsgemeinschaft (SPP „Graphene", TU149/2-1, and Heisenberg Programme, TU149/3-1) and the German Bundesministerium für Bildung und Forschung (BMBF) for financial support. S.K. and U.K. thank the Deutsche Forschungsgemeinschaft and the state Baden-Württemberg for financial support within SALVE (KA 1295/17-2) and (KA 1295/19-1). M.G. and T.W. thank the Deutsche Forschungsgemeinschaft (SPP "Graphene" WE3654/3-1) for financial support.

**Supporting Information**

This material contains the detailed description of techniques and methods, a video file of dark field TEM measurements, model calculations, additional STM, LEED and XPS data.

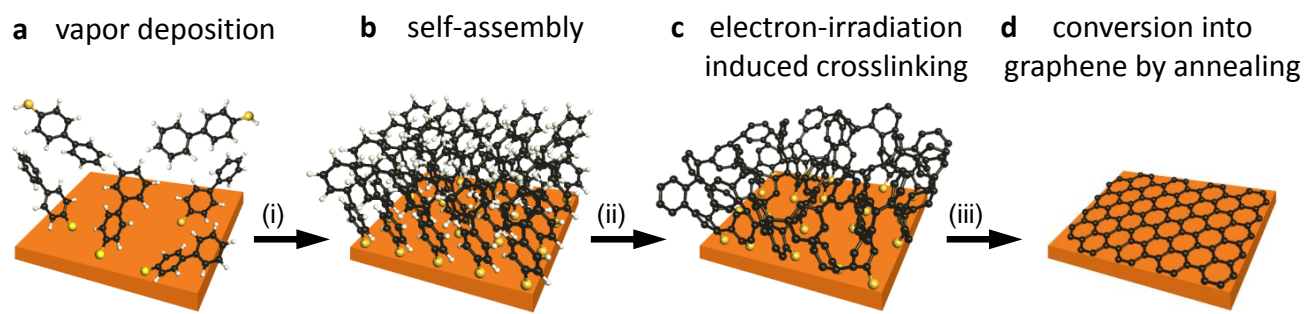

Figure 1

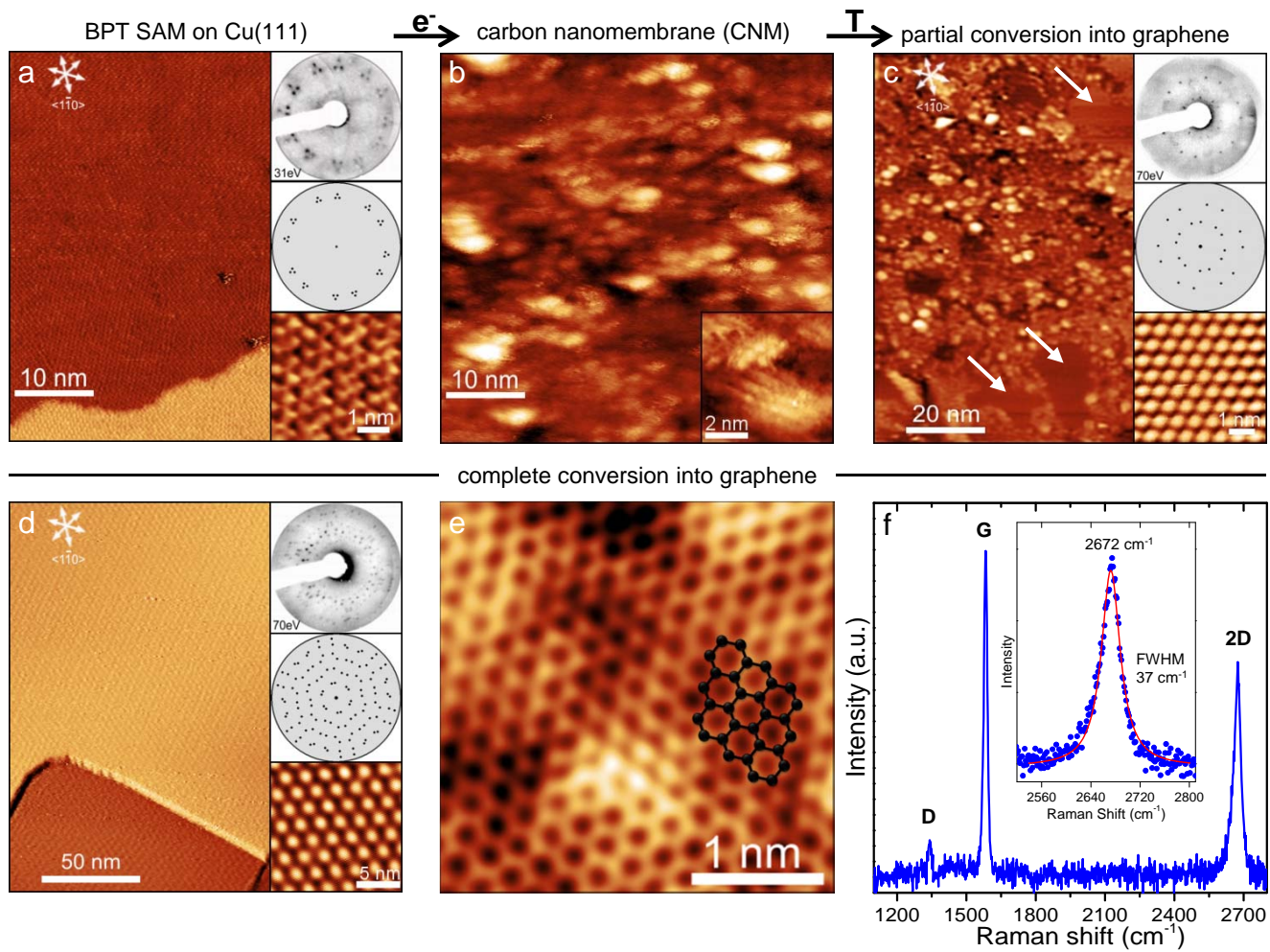

Figure 2

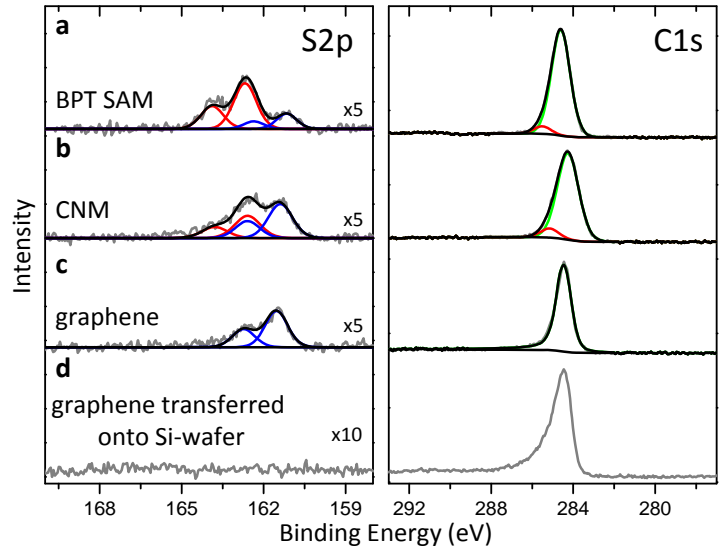

Fig. 3

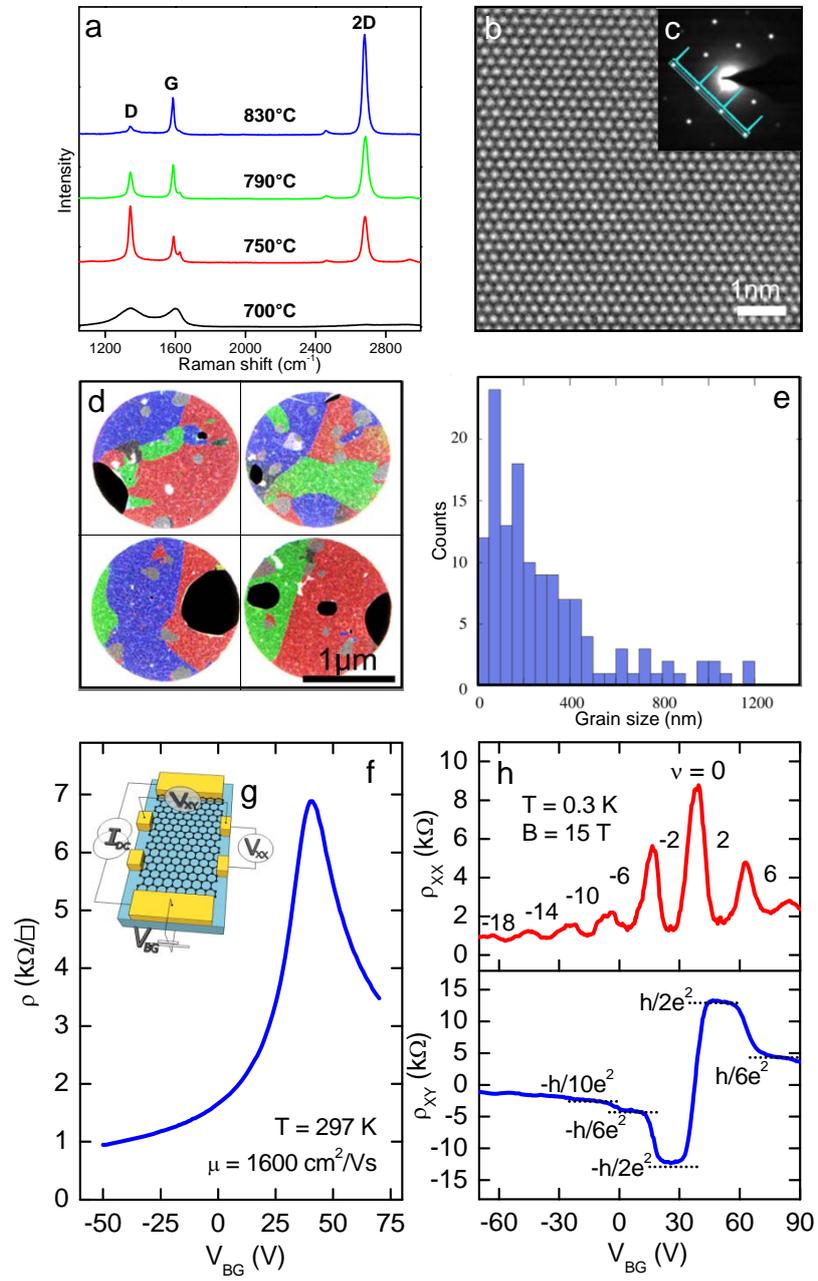

Fig. 4

**Figure captions**

**Figure 1. Schematic of the fabrication route to graphene from aromatic self-assembled monolayers (SAMs) on copper substrates**: a, Deposition of molecules on a substrate; here, vapor deposition of biphenylthiols (BPT) on copper. b, Formation of a SAM. c, Electron-irradiation-induced crosslinking of the BPT SAM into a carbon nanomembrane (CNM). d, Conversion of a CNM into graphene via annealing.

**Figure 2. Characterization by scanning tunneling microscopy (STM) and low energy electron diffraction (LEED) of the conversion of BPT SAMs into graphene on Cu(111).** Simulated LEED patterns are presented below the experimental ones as insets. a, Pristine BPT SAM prepared by vapor deposition on Cu(111) (substrate at RT, evaporation of BPT at 60 °C for 2 h). Lower inset demonstrates a high magnification STM image of one of the structural domains of the BPT SAM. b, The same substrate after electron-irradiation (50 eV) with a dose of 50 mC/cm$^2$ leading to the formation of a CNM. The inset shows a high-resolution STM image of the CNM/Cu substrate. c, Formation of graphene islands within the CNM after UHV annealing for 15 min at 730 °C. The lower inset shows a superstructure of graphene with Cu(111) imaged within these islands. d, Complete conversion of the CNM into graphene after annealing for 2 h at 800 °C (for details see text). The lower inset shows a superstructure of graphene with Cu(111), the atomic structure of graphene is shown in (e). f, Raman spectrum ($\lambda$ = 633 nm) of the formed graphene sheet on Cu(111).

**Figure 3. X-ray photoelectron spectroscopy (XPS) of the conversion of BPT SAMs into graphene on copper substrates.** a, Pristine BPT SAM on Cu(111)

directly after vapor deposition; thiolate and sulfide species are shown in red and blue, respectively. b, The same sample after electron-irradiation (50 eV) with a dose of 50 mC/cm$^2$ leading to the formation of a CNM. c, The sample after annealing for 2 h at 800 °C leading to the conversion into graphene. d, Graphene monolayer prepared on a copper foil and transferred onto a silicon wafer with 300 nm of silicon oxide.

**Figure 4. Spectroscopic, microscopic and electrical characterization of graphene monolayers prepared on copper foils.** a, Raman spectra ($\lambda$ = 532 nm) of the conversion of CNMs into graphene as a function of temperature. The sheets after annealing were transferred from the copper foils onto silicon wafers with 300 nm of silicon oxide. b-e, 80 kV transmission electron microscopy (TEM) of the suspended graphene sheet transferred to a TEM grid after the growth on a copper foil at 850 °C. b, High resolution TEM (HRTEM) micrograph of the sheet clearly resolving the honeycomb lattice of graphene (carbon atoms appear with dark contrast under our imaging conditions (Cs=2 µm, Scherzer defocus)). The single layer nature of this film can be determined already from the HRTEM image contrast; it was further verified by selected area electron diffraction shown in (c), where the intensity ratio between the first and the second order lattice reflections unambiguously identifies the material as single-layer graphene[32]. d, Color coded sequence of dark-field TEM images where different colors correspond to different lattice orientations of graphene crystallites. This method allows to determine the grain size (defined as the square root of the grain area), a corresponding histogram is shown in (e). For more information see supplementary materials. f, Room temperature resistivity of the graphene measured in vacuum as a function of back-gate voltage using Hall bar devices schematically depicted in (g). h, The quantum Hall effect at 0.3 K and 15 T. The upper plot shows Shubnikov-de Haas oscillations with the corresponding filling factors $\nu$ and the lower

plot shows the Hall resistance as a function of back gate voltage, i.e. varied charge carrier density. The measured quantum resistance plateau values are in a perfect agreement with the theoretical sequence for single-layer graphene, $1/N \times h/e^2$ (shown as horizontal dashed lines), where N = ±2, ±6, ±10, …, $h$ and $e$ are the Planck's and the elementary charge constants, respectively.

# SUPPORTING INFORMATION

# Functional single-layer graphene sheets
# from aromatic monolayers

Dan G. Matei[1], Nils-Eike Weber[1], Simon Kurasch[3], Stefan Wundrack[2],
Mirosław Woszczyna[2], Miriam Grothe[2], Thomas Weimann[2], Franz Ahlers[2],
Rainer Stosch[2], Ute Kaiser[3], and Andrey Turchanin[1*]

[1]*Faculty of Physics, University of Bielefeld, 33615 Bielefeld, Germany*
[2]*Physikalisch-Technische Bundesanstalt, 38116 Braunschweig, Germany*
[3]*Electron Microscopy Department of Material Sciences, University of Ulm, Germany*

*e-mail: turchanin@physik.uni-bielefeld.de

**Materials and Methods**

*Growth of BPT SAMs, CNMs and graphene*

1,1'-biphenyl-4-thiol (BPT, H−($C_6H_4$)$_2$−SH) was purchased from Platte Valley Scientific and purified by sublimation. Preparation of BPT self-assembled monolayers (SAMs) on copper substrates was conducted by vapor deposition in an ultra-high vacuum (UHV) preparation chamber integrated into a multi-chamber UHV system (Omicron) with various analytical techniques (see next sections). A Cu(111) single crystal (MaTeck) and polycrystalline copper foils (Alfa Aesar, purity 99.999%, thickness 25 µm), mounted in Mo sample holders, were used as substrates. The copper foils were annealed before use in a tube furnace at 1015 °C for 2 h under a hydrogen atmosphere and a background pressure of 1 mbar to increase their crystallinity. Both the single crystal and the copper foils were *in situ* cleaned before vapor deposition by the Ar$^+$ sputtering (1 keV, 10 mA) at a pressure of 3×10$^{-6}$ mbar for 10 minutes, followed by annealing at 400 °C for 1 h. About 5-6 sputtering and annealing cycles were applied to obtain the atomically clean surfaces for which no carbon was detected by X-ray photoelectron spectroscopy (XPS). Vapor deposition of BPT was conducted with a Knudsen-type evaporator (TCE-BSC, Kentax) from a quartz crucible heated to 50-60 °C. Heating of BPT resulted in an increase of the pressure in the chamber from ~$10^{-10}$ mbar to ~$10^{-7}$ mbar, as detected by $N_2$-calibrated vacuum ion gauge. The copper substrates were kept at room temperature (RT) during the vapor deposition. Typical evaporation time was between 1 and 2 h. The formed BPT SAMs were then cross-linked into carbon nanomembranes (CNMs) under UHV conditions in an analysis chamber of the same multi-technique UHV system (Omicron) under a background pressure of $5*10^{-10}$ mbar using an electron flood-gun (SL1000, Omicron) at an energy of 50 eV and a dose of 50 mC/cm². Annealing of the CNMs leading to the conversion into graphene was conducted on a heatable/coolable manipulator with a PBN resistive heater placed below the sample. Temperature of the samples was controlled on both copper sample and molybdenum sample holders areas employing a two-color pyrometer (SensorTherm) with the emissivity coefficients of 5 % and 23 %, respectively. Heating and cooling (liquid nitrogen) of the samples to the target temperatures was achieved in about 15 min.

*Transfer of as-grown graphene sheets*

As-grown graphene sheets on copper foils were transferred onto silicon wafers with oxidized surface layer (300 nm) and TEM-grids (Plano S147-3 and Quantifoil R1.2/1.3 on Au 200 mesh) using the following procedure. Firstly, a poly(methyl methacrylate) (PMMA, AR-P 631.04) layer was spincast onto the surface for 30 s at 4000 rpm; next, the graphene/PMMA sandwich was baked on a hotplate at 90°C for 5 min and then a second PMMA (AR-P 671.04) layer was spincast and baked. After removing by O$_2$-plasma (2 min) the graphene formed on the copper back-side of this sandwich, the copper was etched away in a 0.3 M ammonium persulfate solution (>98%, Sigma Aldrich) for ~15 hours. After cleaning in water for five minutes the PMMA/graphene sandwich was then transferred onto a new target substrate. Then the sandwich was dried with a gentle flow of nitrogen and annealed at 90 °C for 5 min. The resist was removed by immersion in acetone for 1 h. The graphene on silicon wafer samples were afterwards dipped in isopropanol and blown dry with nitrogen. Graphene samples on TEM-grids were treated in a critical point dryer (Tousimis, Autosamdri-815) to minimize damage of the freestanding parts.

*Fabrication of Hall bar structures*

Graphene Hall bars on Si/SiO$_2$-wafers (As-doped, resistivity 3-7 mΩcm, with 300 nm of the thermally grown silicon oxide, Si(100)) were fabricated using standard electron beam lithography and PMMA masks. Geometrical definition of the graphene shapes was achieved by dry etching in argon/oxygen plasma. Ti/Au contacts (10 nm/100 nm) were made by thermal evaporation and lift-off. The dimensions of Hall bars were 18 μm by 7 μm with a distance between side contacts of 5 μm. The fabricated devices were mounted onto 10 mm by 10 mm printed circuit boards using silver epoxy glue and gold wires were bonded to form electrical connections.

*X-ray photoelectron spectroscopy (XPS), low energy electron diffraction (LEED) and scanning tunneling microscopy (STM)*

A multi-chamber UHV-system (Omicron) consisting of an analysis chamber equipped with STM (Multiscan VT), LEED and XPS was employed for the *in situ* analysis of the samples. X-ray photoelectron spectra were recorded using a monochromatic X-ray source (Al K$_\alpha$) and an electron analyzer (Sphera) with a resolution of 0.9 eV. The effective thickness of the monolayers was estimated from the exponential attenuation of the substrate Cu 2p$_{3/2}$ signal in comparison to the signal of a clean Cu(111) reference using an attenuation length of 18 Å. Binding energies were calibrated with respect to the Cu 2p$_{3/2}$ signal at 932.6 eV. For peak fitting a Shirley background and Voigt functions were used. STM imaging was conducted with a Multiscan VT microscope using electro-chemically etched tungsten tips with tunneling currents of 30-80 pA and bias voltages of 300 mV. LEED patterns were recorded using a BDL600IR-MCP (OCI Vacuum Microengineering) system with a multi-channel plate detector. Experimental LEED patterns were simulated with the LEEDsim software. Following the short exposures to electron or X-ray beams from LEED and XPS instruments, respectively, no changes were detected in the samples by STM.

*Raman Spectroscopy*

Raman spectra were acquired using a micro Raman spectrometer (LabRam ARAMIS) operated in the backscattering mode. Measurements at 532 and 633 nm were obtained with a frequency-doubled Nd:YAG-Laser and a HeNe Laser, a 100x objective and a thermoelectrically cooled CCD detector (2-3 cm$^{-1}$ spectral resolution). The Si-peak at 520.5 cm$^{-1}$ was used for peak shift calibration of the instrument.

*Transmission electron microscopy (TEM)*

All TEM experiments were carried out in an aberration corrected FEI Titan 80-300 microscope operated at 80 kV. The extraction was reduced to 2000 V to minimize the energy spread of the gun. The high resolution TEM data were recorded at Scherzer conditions with Cs ~ 2 μm and an underfocus of about -10 nm. The DF sequence was recorded with the condenser aperture placed on the optical axes and an incident beam tilt of ~1°. Then the beam was rotated at constant tilt from 0 to 60° in 5° steps and subsequent dark field (DF) images were recorded. Most important, for the color coded DF images shown in Figure 4d of the main article we used only 3 frames of the DF sequence where the grain size distribution was extracted out of the complete dataset that can be seen in the supplementary video. For the evaluation of the grain sizes, the average signal of the sequence was subtracted from the individual slides to remove the contribution of the supporting carbon film and bring out the grains underneath as can be seen in the top row of SI Figure 5. The lower row shows electron diffraction patterns obtained from the areas indicated by the colored ring. As can be seen from the intensity distribution, both of these areas are built of single layer single crystal graphene.

*Electrical transport measurements*

Electrical transport measurements were carried out in an Oxford Instruments Helium 3 refrigerator HelioxTL. Before cooling down to the base temperature of 0.3 K each sample was kept in vacuum at $10^{-6}$ mbar for 18 hours and subsequently electrically characterized. Keithley 2400 SourceMeter instruments were utilized to apply direct current (1 µA in all electrical measurements) and back-gate voltage. Voltages were measured by Keithley 2182A Nanovoltmeters.

**Video file**

Complete dataset of the DF-TEM images sequence used for the evaluation of the grain size distribution.

**Evaluation of the number of graphene layers formed by the conversion of a BPT SAM on Cu(111)**

Area for the unit cell for a BPT SAM on Cu (111) is given by:

$$A_{BPT} = a_1 * a_2 * \sin(\alpha_2) = 5.35 \text{Å} * 5 \text{Å} * \frac{\sqrt{3}}{2} = 22.56 \text{ Å}^2.$$

Area for the unit cell of graphene is given by:

$$A_{graphene} = a_{gr} * a_{gr} * \sin(\alpha_2) = 2.46 \text{Å} * 2.46 \text{Å} * \frac{\sqrt{3}}{2} = 5.24 \text{ Å}^2$$

Taking into account the number of carbon atom per unit cell in the BPT SAM ($n_{BPT}$=12) and in the monolayer of graphene ($n_{graphene}$= 2) the number of graphene layers ($N$) after the complete conversion of a BPT SAM can be estimated as:

$$N = \frac{A_{graphene} * n_{BPT}}{n_{graphene} * A_{BPT}} = \frac{5.24 * 12}{2 * 22.56} \approx 1.4$$

Correcting this number for carbon desorption during the crosslinking and annealing, which is in total ~30% ($f_{XPS}$), as detected by XPS, we obtain for the number of graphene layers:

$$N_{corrected} = N - (f_{XPS} * N) = 1.4 - (0.3 * 1.4) \approx 1.$$

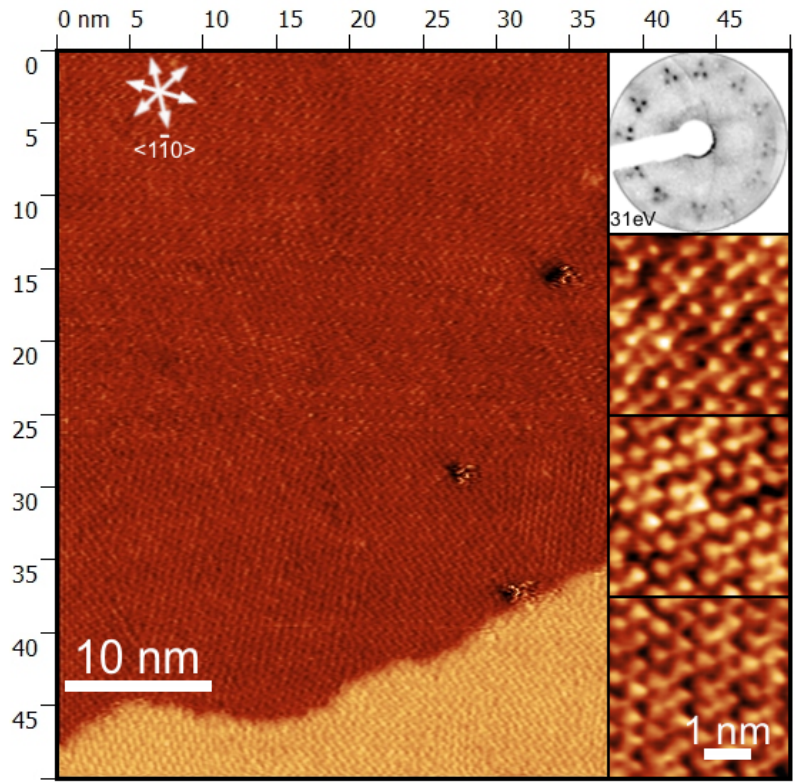

**SI Figure 1│ High-resolution STM images of a BPT SAM obtained by UHV evaporation on a Cu(111) surface.** The top inset shows a LEED pattern obtained from the sample that shows an incommensurate structure of the SAM with respect to the substrate with several rotational domains. Three of these domains are presented in the lower insets.

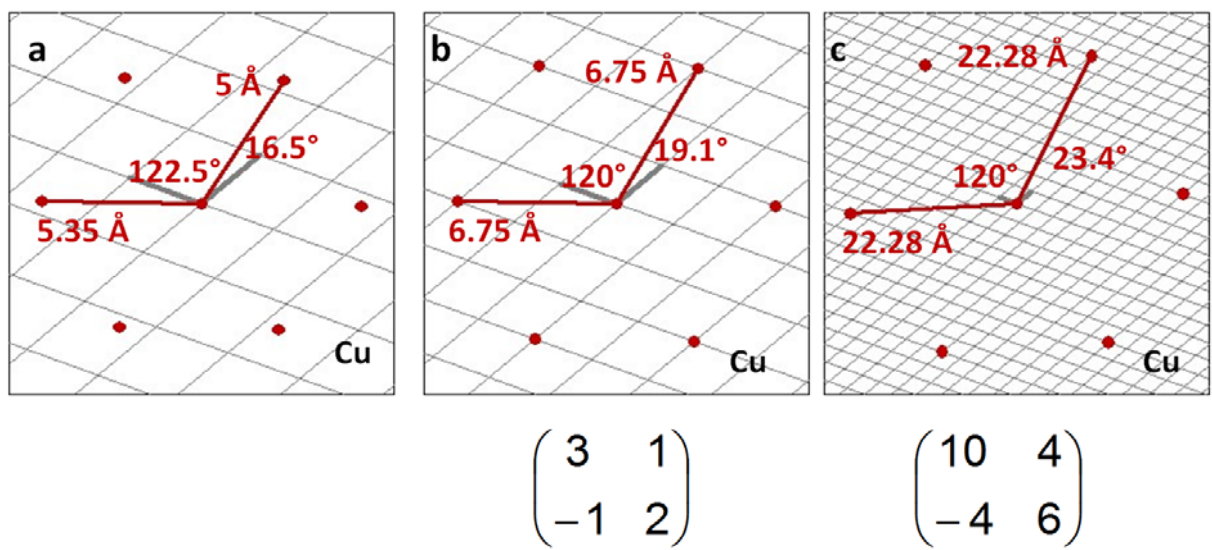

**SI Figure 2 | Schematic drawings of the unit cells used for the simulation of LEED patterns: a**, structure of the BPT SAM on Cu(111); **b**, superstructure of graphene on Cu(111) shown in the lower inset of Figure 2c; **c**, superstructure of graphene on Cu(111) shown in the lower inset of Figure 2d. The structures in **b** and **c** are commensurate with the Cu(111) substrate; their respective matrices are presented below the drawings.

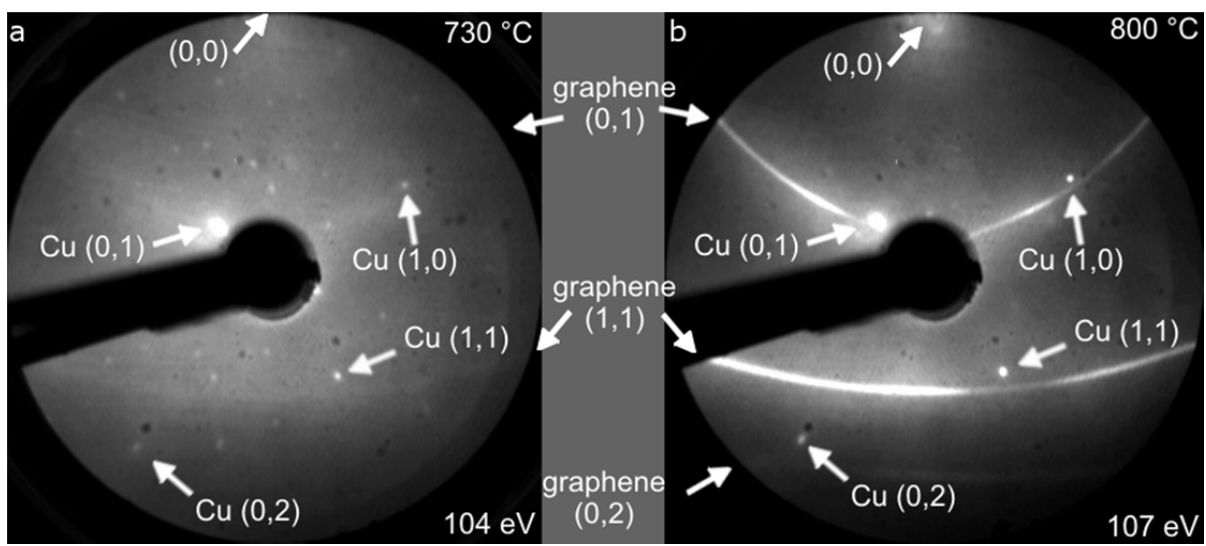

**SI Figure 3 │ Tilted LEED patterns of the samples annealed at 730 °C (see Fig. 2c) and 800 °C (see Fig. 2d) presenting the formation of randomly oriented graphene crystallites (see diffraction rings).** Different diffraction features of Cu(111) and graphene are indicated within the images. For more details see main paper.

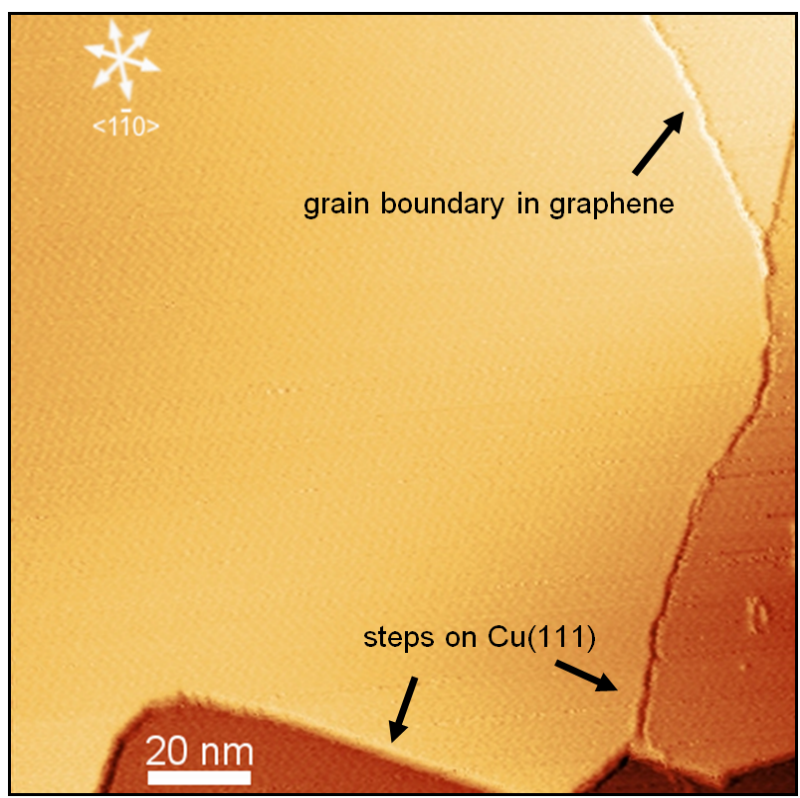

**SI Figure 4**│ **Large-area STM image of graphene on Cu(111) obtained after annealing the cross-linked BPT SAM (CNM) at 800 °C for 2 h**.

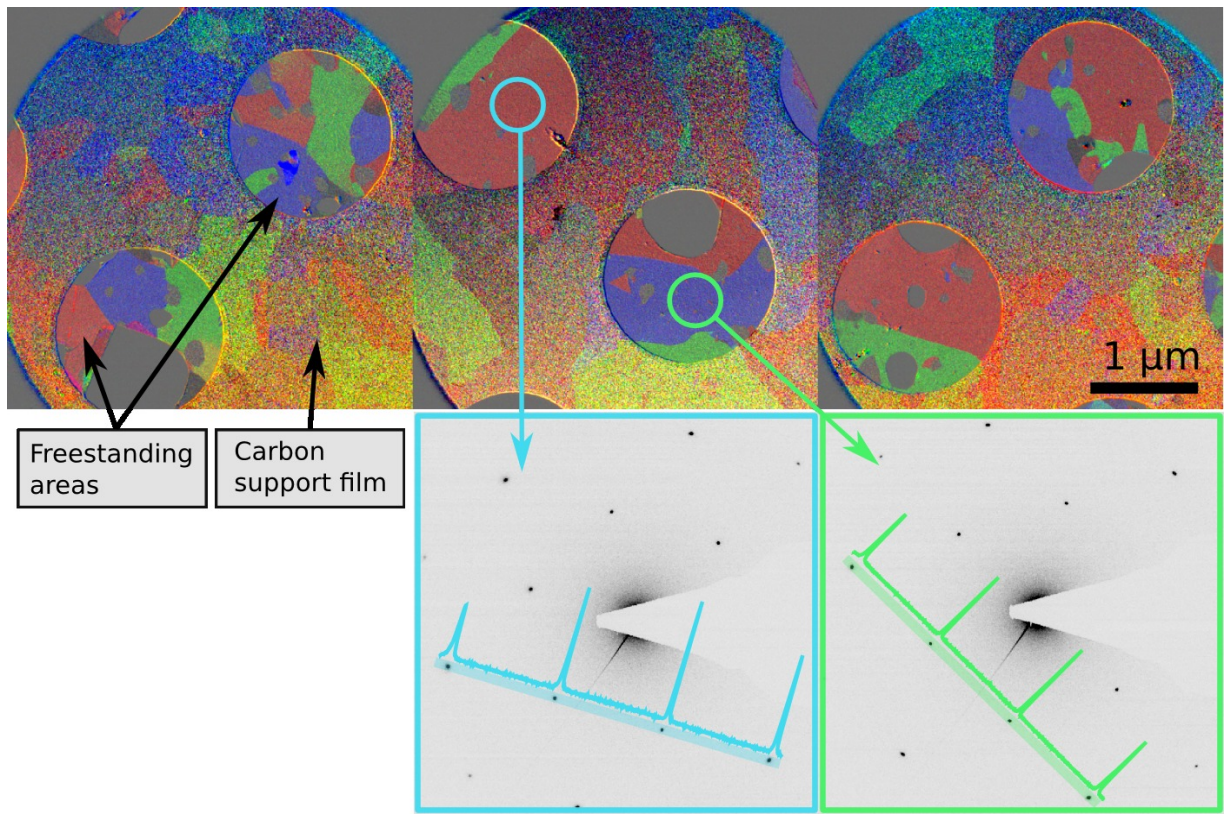

**SI Figure 5 | Characterization of the polycrystallinity of graphene monolayers prepared on copper foils.** The top row shows a colored sequence of dark field images. The average signal was removed to see the grain structure of graphene sheets on top of the supporting holey amorphous carbon film. The lower row shows selected area electron diffraction patterns from the areas indicated by the colored circles.

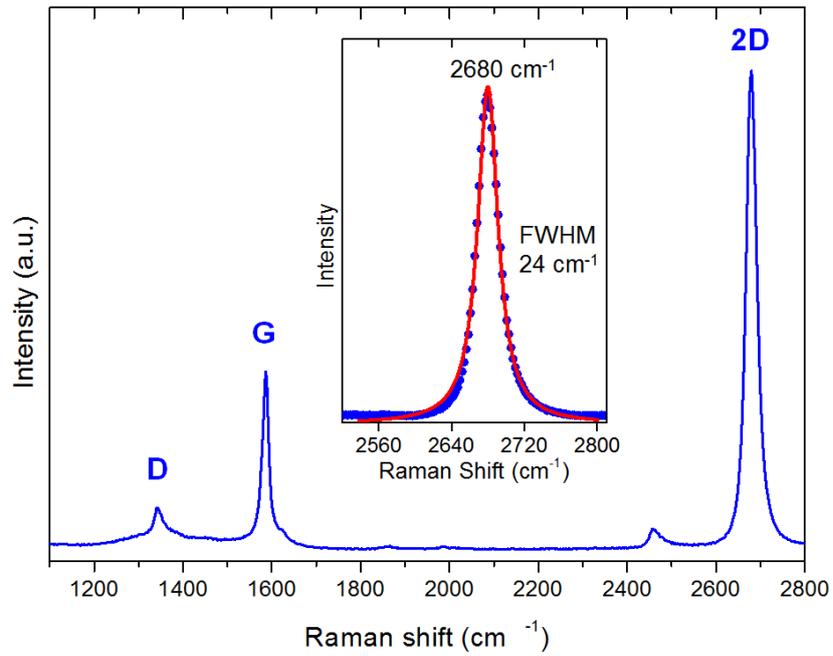

**SI Figure 6│ Raman spectrum (λ = 532 nm) of a graphene sheet prepared on a copper foil (830 °C) and then transferred onto a silicon wafer with 300 nm of silicon oxide.** Insert is a Lorentzian fit of the 2D peak.

**SI Table 1 | Experimental X-ray photoelectron spectroscopy data obtained for BPT SAMs, CNMs and graphene on Cu(111).** For details see main paper.

| C1s | Binding energy, eV | FWHM, eV | Area, % | element loss, % |
|---|---|---|---|---|
| SAM | | | | 0 |
| C-C | 284.6 | 1.1 | 94 | - |
| C-S | 285.5 | 1 | 6 | - |
| CNM | | | | 5 |
| C-C | 284.3 | 1.2 | 92 | - |
| C-S | 285.2 | 1.1 | 8 | - |
| graphene | | | | 30 |
| C-C | 284.5 | 0.9 | 100 | - |
| S2p$_{3/2}$ | | | | - |
| SAM | | | | 0 |
| thiolate | 162.7 | 1 | 77 | - |
| copper sulfide | 161.2 | 0.9 | 23 | - |
| CNM | | | | 0 |
| thiolate | 162.6 | 1.1 | 41 | - |
| copper sulfide | 161.4 | 1 | 59 | - |
| graphene | | | | 40 |
| copper sulfide | 161.5 | 1 | 100 | - |